\documentclass[aps,prb,twocolumn,superscriptaddress,amsmath]{revtex4-2}

\usepackage[colorlinks=true, linkcolor=blue, filecolor=blue, urlcolor=blue, citecolor=blue]{hyperref}
\usepackage{amsfonts}
\usepackage{bbm}
\usepackage{graphicx}
\usepackage{dcolumn}
\usepackage{bm}
\usepackage{soul}

\begin{document}

\title{Tunable Majorana corner states driven by superconducting phase bias in a vertical Josephson junction}

\author{Cheng-Ming Miao}
\email[]{These authors contribute equally to this work.}
\affiliation{International Center for Quantum Materials, School of Physics, Peking University, Beijing 100871, China}

\author{Yu-Hao Wan}
\email[]{These authors contribute equally to this work.}
\affiliation{International Center for Quantum Materials, School of Physics, Peking University, Beijing 100871, China}

\author{Ying-Tao Zhang}
\affiliation{College of Physics, Hebei Normal University, Shijiazhuang 050024, China}

\author{Qing-Feng Sun}
\email[]{sunqf@pku.edu.cn}
\affiliation{International Center for Quantum Materials, School of Physics, Peking University, Beijing 100871, China}
\affiliation{Hefei National Laboratory, Hefei 230088, China}

\begin{abstract}
The realization and manipulation of Majorana zero modes is a key step in achieving topological quantum computation.
In this paper, we demonstrate the existence of Majorana corner states in a superconductor-insulators-superconductor vertical Josephson junction.
The position of these Majorana corner states can be precisely and easily controlled by the superconducting phase bias, which be confirmed through both numerical and edge state theoretical analysis.
In addition, we propose a protocol for achieving topological braiding of the Majorana corner states in a system of three circular vertical Josephson junctions. Our findings advance the field of topological quantum computation by providing new insights into the efficient and precise manipulation of Majorana corner states.
\end{abstract}

\maketitle

\section{\label{sec1}Introduction}
Majorana zero modes are exotic quasiparticles that emerge as zero-energy solutions in the Majorana equation~\cite{Majorana1937}. They obey non-Abelian exchange statistics and resist local perturbations, making them promising for topological quantum computation \cite{Dennis2002, Nayak2008, Kitaev2001, Ivanov2001, Alicea2012, Leijnse2012, Nadj2013, Beenakker2013,Yazdani2023,Kouwenhoven2025}. In condensed matter physics, Majorana zero modes are predicted to appear at boundaries or defects of topological superconductors (SCs), with platforms such as quantum wires~\cite{Lutchyn2010, Oreg2010, Mourik2012, Das2012, Rokhinson2012, Finck2013, Churchill2013, Miao2022, Zhang2024}, superconducting vortices \cite{Hosur2011, Xu2016, Wang2015, Hell2017, Zhang2018}, Josephson junctions~\cite{Fornieri2019, Haim2019, Hart2017, Pientka2017, Ren2019, Setiawan2019, Stern2019}, and SC-proximitized topological insulators or quantum anomalous Hall insulators (QAHIs)~\cite{Fu2008, Qi2010, Chung2011, He2014, Zhang2017}. Observing their exchange statistics is crucial for topological braiding.
To this end, various manipulation approaches have been proposed including electrostatically tuned gate voltages~\cite{Alicea2011, Karzig2016, Yan2019a, Zhou2019, Hodge2025}, strain-induced state density variation~\cite{Wang2022, Li2022}, and probe-mediated vortex movement~\cite{Posske2020, Ma2020}. 
Furthermore, two-dimensional second-order topological SCs host Majorana corner states (MCSs), special Majorana zero modes, offering inherent advantages for braiding~\cite{Liu2018, Wang2018, Yan2018, Zhu2018, Zhu2019, Yan2019, Laubscher2020, Wu2020, Tan2022, Pan2024}. 
Unlike Majorana zero modes in rigorous one-dimensional systems, MCSs can be exchanged without mixing, simplifying the braiding process. Based on this, several braiding schemes have been proposed including in-plane magnetic field tuning \cite{Zhang2020, Pahomi2020, Ikegaya2021, Liu2024, Lapa2021}, LC-circuit emulation \cite{Ezawa2019}, electrically controlled chemical potentials \cite{Zhang2020a}, among others.
Magnetic-field-based approaches \cite{Zhang2020, Pahomi2020, Ikegaya2021, Liu2024, Lapa2021} face significant challenges due to their reliance on high-precision magnetic fields (and sometimes
simultaneous phase control) leads to imprecise operations or complex implementation. Meanwhile, existing magnetic-field-free methods \cite{Ezawa2019, Zhang2020a} exhibit their own limitations, such as 
limited mobility and flexibility of Majorana zero modes hinder efficient and continuous braiding operations \cite{Ezawa2019}, as well as sensitivity to edge geometry and imperfections may disturb the braiding evolution~\cite{Zhang2020a}.
Building upon these insights, our work aims to develop a manipulation scheme that is efficient, precise, continuous, robust, and easily implementable. Notably, overcoming these challenges will make a significant contribution to the realization of topological quantum computing.

\begin{figure}
  \centering
  \includegraphics[width=\columnwidth,angle=0]{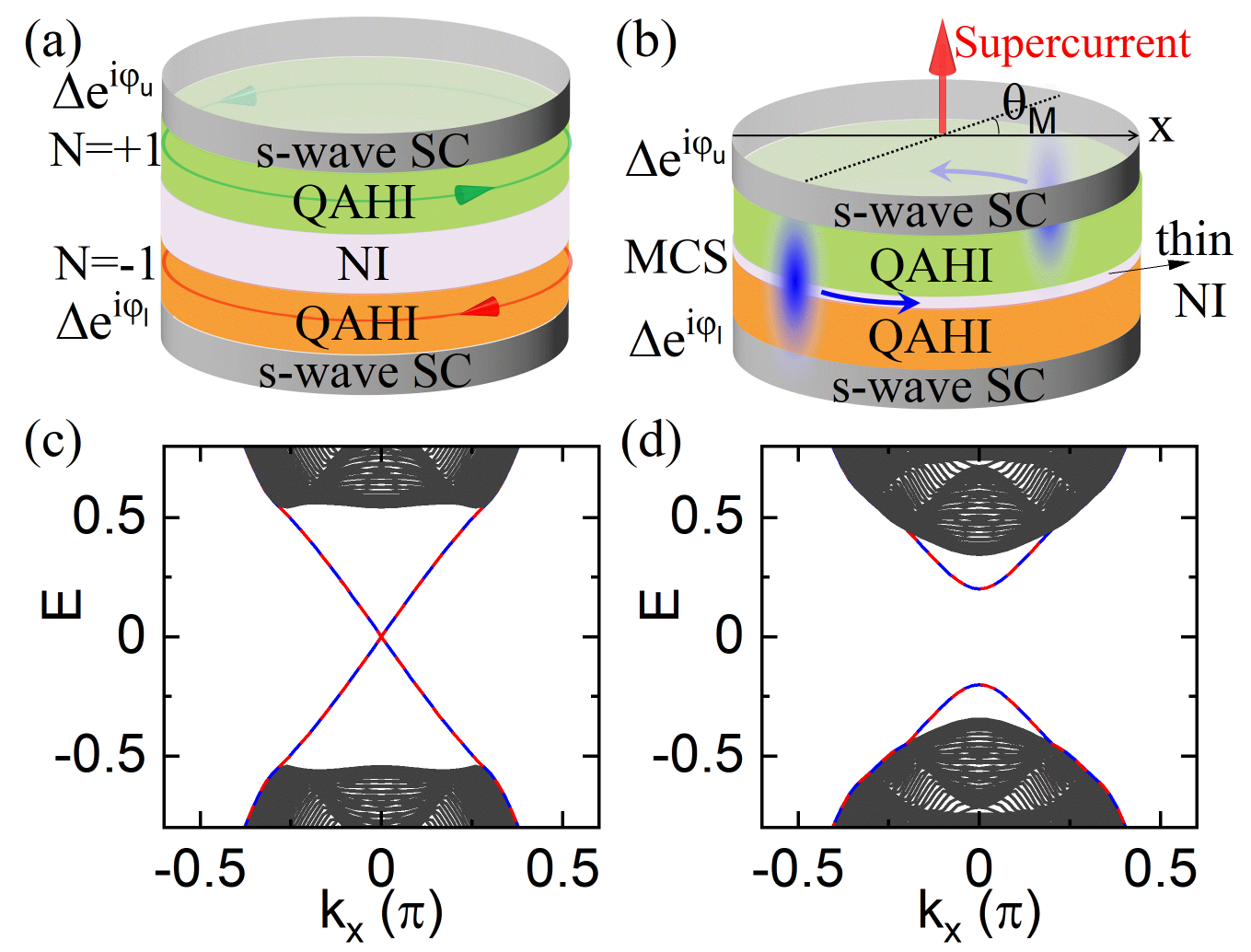}
  \caption{(a,b) Schematic of the SC/QAHI/NI/QAHI/SC vertical Josephson junction. The green and orange QAHIs host opposite Chern numbers. $\Delta$ and $\varphi_{u(l)}$ represent the superconducting pairing potential and phase of the grey $s$-wave SC.
(a) Majorana edge states propagate counterclockwise/clockwise in the upper/lower layers (green/red lines with arrows). (b) A thin NI layer introduces non-zero interlayer coupling, disrupting the two Majorana edge states and generating two tunable MCSs (in blue), which are driven by supercurrent (red arrow).
$\theta_M$ is the angle between the line connecting two MCSs and the $+x$ axis.
(c,d) Band structures of nanoribbons with $x$-direction periodic boundary conditions for different interlayer coupling strengths $t = 0$ in (c) and $t = 0.2$ in (d). The alternating red and blue lines represent Majorana edge states. The other parameters are set to be $m=-0.5$, $\mu=1$, $A=1$, $B=1$, $\Delta=1$, $\varphi_u=0$, $\varphi_l=\varphi=0$, and $N_{y}=100a$.}
  \label{fig1}
\end{figure}

In this paper, we propose a scheme for manipulating MCSs by superconducting phase bias in an SC-middle insulators-SC vertical Josephson junction without a magnetic field. Specifically, the middle insulators are structured as a QAHI-normal insulator (NI)-QAHI sandwich structure, and the superconducting phase bias can be controlled by a supercurrent [see Figs. \ref{fig1}(a,b)]. It is demonstrated that superconducting phase bias can effectively manipulate the position of MCSs.
The emergence and movement of these MCSs can be explained by the edge states orthogonal theory.
In addition, we propose a topological braiding protocol by exchanging the positions of MCSs in a hybrid system of three circular vertical Josephson junctions (CVJJs). Our findings advance topological quantum computation by offering new insights into the efficient, precise, continuous, robust, and easy manipulation of MCSs by superconducting phase bias.

The paper is structured as follows.
In Sec. \ref{sec2}, we present the model Hamiltonian for the  CVJJs.
Section \ref{sec3} demonstrates the emergence of MCSs and shows that two such MCSs can be efficiently and easily manipulated by tuning the superconducting phase bias.
In Sec. \ref{sec4}, we elaborate on the origin of MCSs, and prove that their positions can be precisely and continuously controlled based on numerical and analytical results.
Then in Sec. \ref{sec5}, we propose a protocol for achieving the topological braiding by swapping two of the four MCSs.
Finally, a summary is presented in Sec. \ref{sec6}.
Some auxiliary materials are relegated to Appendices \ref{appendixA} to \ref{appendixC}.

\section{\label{sec2}model and hamiltonian}

We start from the QAHI-NI-QAHI sandwich junction, which has been experimentally reported to achieve multiple chiral edge states~\cite{Zhao2020}. It is worth noting that the two QAHIs here host opposite Chern numbers, which can be achieved in a variety of ways by setting opposite magnetic orderings, applying strain, and twisting the materials, etc.~\cite{Xu2023, Guo2023, Xue2024, Li2024}.
When the intermediate NI layer is sufficiently thick, the interlayer coupling between the upper and lower QAHI layers can be ignored. In this case, we consider the simplest model Hamiltonian for upper/lower QAHI layer~\cite{Qi2006}:
\begin{align}   h_{\mathbf{k}}^{u/l}=\left( m-B\mathbf{k}^2 \right) \sigma _z \pm A\sigma _xk_x+Ak_y\sigma _y,
  \label{eq1}
\end{align}
where $\mathbf{k} = (k_x, k_y)$ is a wave vector in the first Brillouin zone. $m$, $A$ and $B$ are material parameters, and $\sigma_{x,y,z}$ are the Pauli matrices acting on the spin space. Here, the second term in the right side of Eq. (\ref{eq1}) is respectively $+$ and $-$ for the upper and lower QAHI layer, which corresponds to their Chern numbers being $+1$ and $-1$.

In proximity to the $s$-wave SC, a finite pairing potential $\Delta$ can be induced in QAHI. The Bogoliubov-de Gennes (BdG) Hamiltonian for the proximity-coupled QAHI is:
\begin{align}
 H_{\mathbf{k}}^{u/l}&=\left(\begin{array}{cc}
			h_{\mathbf{k}}^{u/l}+\mu & i\Delta e^{-i\varphi_{u/l}}\sigma_{y}\\
			(i\Delta e^{-i\varphi_{u/l}} \sigma_{y})^{\dagger}   & -h_{\mathbf{-k}}^{u/l*}-\mu
		\end{array}\right),
  \label{eq2}
\end{align}
where $\mu$ is the chemical potential.
$\varphi_{u/l}$ is the superconducting phases of the upper/lower $s$-wave SC.
The nontrivial topological phase with Chern number $\mathcal{N} = +1$/$\mathcal{N} = -1$ of the upper/lower layer is obtained by setting $m^2<\Delta^2+\mu^{2}$, which supports chiral Majorana edge states (MESs) propagating counterclockwise/clockwise in the upper/lower layers~\cite{Qi2010, Chung2011, He2014, Zhang2017}, as shown in Fig. \ref{fig1}(a).

While the intermediate NI is thin, the two converse chiral MESs are destroyed due to the introduction of interlayer coupling [see Fig. \ref{fig1}(b)].
Then the Hamiltonian for the SC/QAHI/NI/QAHI/SC vertical junction is:
\begin{align}
  H_{\mathbf{k}}=\left(\begin{array}{cc}
			 H_{\mathbf{k}}^{u} & t\tau_{0} \sigma_{0}\\
			(t \tau_{0}\sigma_{0})^{\dagger}   &  H_{\mathbf{k}}^{l}
		\end{array}\right),
  \label{eq3}
 \end{align}
where $t$ represents the strength of the interlayer coupling, which is related to the thickness of the NI layer.
$\tau_{0}$ and $\sigma_{0}$ denote the $2\times 2$ identity matrix acting on Nambu and spin spaces, respectively. The tight-binding form with square lattices of the above continuum model is shown in Appendix \ref{appendixA}.
In our calculations, we set $A=B=1$, $m=-0.5$ and $\Delta=1$, which ensures that the system is in a nontrivial topological phase, as the condition $m^2<\Delta^2+\mu^{2}$ is satisfied.
We focus on the effects of the superconducting phase bias, with the parameters fixed as $\varphi_{u}=0$ and $\varphi_{l}=\varphi$.  In fact, the results only depend on $\varphi_{l}-\varphi_{u}$.

\section{\label{sec3} The realization and manipulation of two MCSs}

To confirm the MESs analyzed above, we plot the energy bands of the vertical junction nanoribbons with $x$-direction periodic boundary conditions for different interlayer coupling strengths in Figs.~\ref{fig1}(c,d).
Here, we choose the nanoribbon width to be $N_y = 100a$, where $a=1$ is the lattice constant.
Figure \ref{fig1}(c) shows the presence of two gapless chiral MESs in the Josephson junction with $t=0$, corresponding to $\mathcal{N}=+1/\mathcal{N}=-1$ in the upper/lower layer. In the presence of the interlayer coupling $t=0.2$, the MESs are gapped, as illustrated by alternating red and blue lines in  Fig. \ref{fig1}(d).

\begin{figure}
	\centering
	\includegraphics[width=\columnwidth,angle=0]{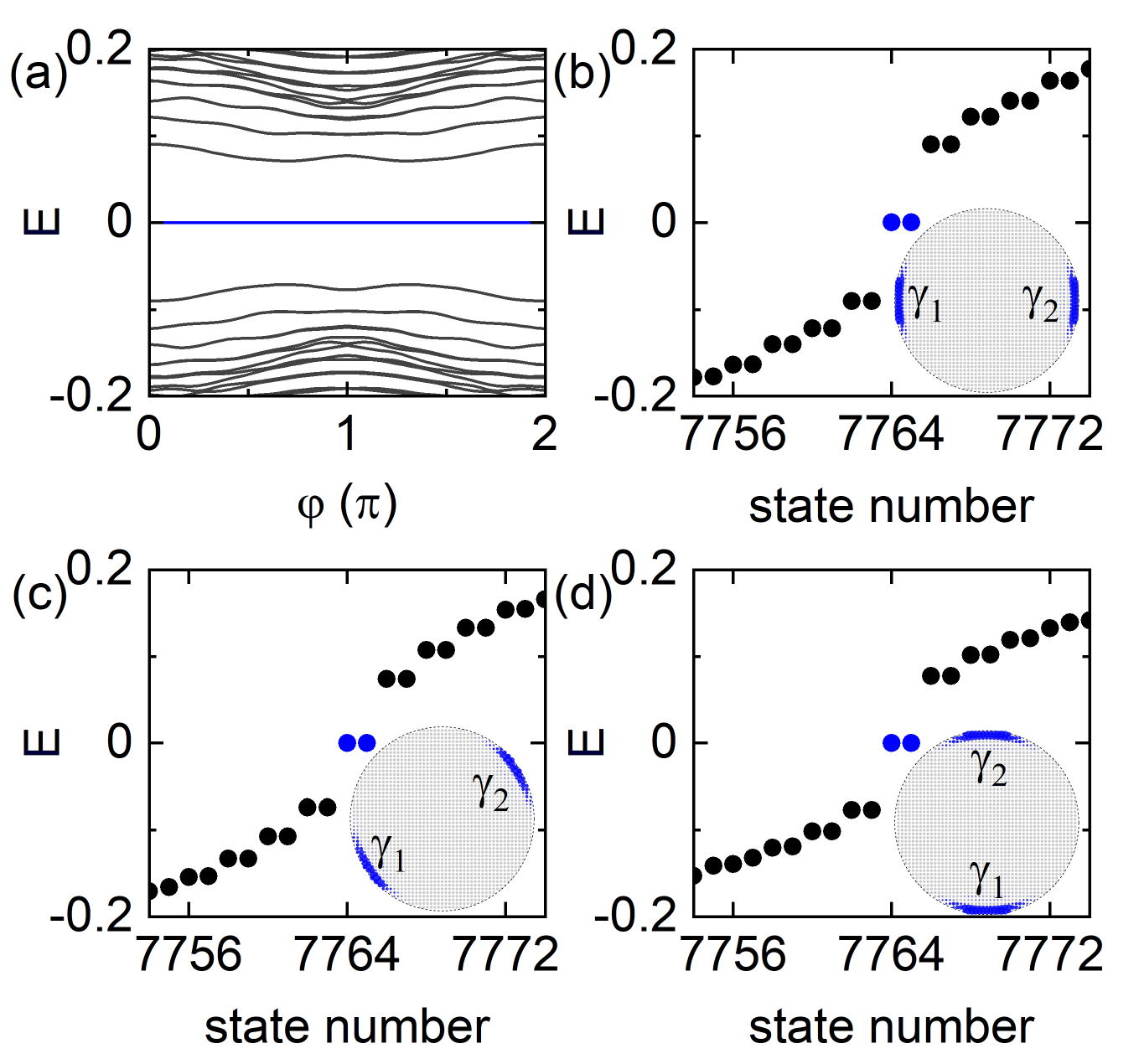}
	\caption{(a) Band structure for the CVJJ as a function of the superconducting phase bias $\varphi$. Blue lines indicate the Majorana zero-energy bands, which always remain separate from the other bands (black lines). (b)-(d) Energy levels of the CVJJ with different superconducting phase biases $\varphi=0$ for (b), $\varphi=0.5\pi$ for (c) and $\varphi=\pi$ for (d). The blue dots represent the in-gap MCSs, the probability distribution of the MCSs is plotted in the inset. The radius is set to $R=25a$ and other parameters are the same as those in Fig. \ref{fig1}(d).}
	\label{fig2}
\end{figure}

To explore the second-order topology, we plot the energy levels for the CVJJ versus superconducting phase bias $\varphi$ in Fig. \ref{fig2}(a).
The geometry of the CVJJ is set as: we construct a square lattice plane centered at the origin $(0,0)$, where the position of each site is labeled by its coordinates $(x,y)$. The sites belonging to the CVJJ satisfy the condition $x^{2}+y^{2} < R^{2}$ with radius $R=25a$.
It is evident that the zero-energy in-gap states (blue lines) remain stable and separate from the
other bands (black lines) throughout the variation of $\varphi$ [see Fig. \ref{fig2}(a)].
Moreover, we plot the distribution of these in-gap states with different $\varphi$ in Figs. \ref{fig2}(b-d).
As shown in Fig. \ref{fig2}(b), the in-gap Majorana states localize at the left and right corners of the CVJJ with $\varphi=0$. Figure \ref{fig2}(c) shows that the Majorana states are bound at the upper-right and lower-left corners of the CVJJ with $\varphi=0.5\pi$.
As the $\varphi$ increases to $\pi$, the position of the MCSs shift to the upper and lower corners in Fig. \ref{fig2}(d).
These results intuitively demonstrate that MCSs occur in the SC/QAHI/NI/QAHI/SC vertical Josephson junction and their positions can be precisely and easily controlled by the magnitude of the superconducting phase bias $\varphi$.  Furthermore, this manipulation remains robust in the elliptic vertical Josephson junctions, indicating a degree of insensitivity to geometric shape (see Appendix \ref{appendixB}).

\section{\label{sec4}The origin of the MCSs}

To gain a thorough understanding of the MCSs, we analyze the chiral MESs in the $s$-wave SC proximity-coupled QAHI. Combining Eq. (\ref{eq1}) with Eq. (\ref{eq2}), the low-energy Hamiltonian can be expressed:
\begin{align}
  H_{\mathbf{k}}^{u/l}=&(m-B\mathbf{k}^2)\tau_{z}\sigma_{z}\pm Ak_{x}\tau_{0}\sigma_{x}+Ak_{y}\tau_{z}\sigma_{y} +\mu\tau_{z}\sigma_{0}\nonumber \\
  &-\Delta\cos{\varphi_{u/l}}\tau_{y}\sigma_{y}+\Delta\sin{\varphi_{u/l}}\tau_{x}\sigma_{y},
  \label{eq4}
 \end{align}
where $\tau_{x,y,z}$ are the Pauli matrices acting on Nambu space.
We consider a circular-shaped boundary, the normal direction of the boundary tangent for arbitrary angle $\theta$ is $\hat{x}_{\bot}= (\cos \theta, \sin \theta)$.
Next, we assume an ansatz for the edge state wave function at $\theta$ as $\Psi_{u/l}(x_{\bot})=e^{\lambda x_{\bot}}e^{i k_{\parallel} x_{\parallel}} \left|\xi_{u/l}\right>$ with $k_\parallel=\sin{\theta}k_{x}-\cos{\theta}k_{y}$.
Here, $\left|\xi_{u}\right>$ and
$\left|\xi_{l}\right>$ represent the spinor components of the chiral MESs in the upper and lower layers, respectively, and they determine whether two MESs are orthogonal.
Considering a half-infinite sample area $x_{\bot} < 0$, we can solve out the spinors of the chiral MESs with $\varphi_u=0$ and $\varphi_l=\varphi$:
\begin{align}
\small
   \left|\xi_u\right>=\left[\begin{matrix}
		\xi_1 \\
		\xi_2/[i  e^{-i\theta} (\lambda-k_\parallel) ]\\
	\xi_3/[i  e^{-i\theta} (\lambda-k_\parallel) ] \\
		\xi_4
	\end{matrix}\right] ,
   \left|\xi_l\right>=\left[\begin{matrix}
		\xi_1 \\
		\xi_2/[-i  e^{i\theta} (\lambda+k_\parallel) ]\\
	\xi_3/[-i  e^{i(\theta-\varphi)} (\lambda+k_\parallel) ] \\
		\xi_4/[e^{-i\varphi}]
	\end{matrix}\right] ,
    \label{eq5}
\end{align}
where $\xi_1=-\Delta^2-(m+\Lambda)^2+\mu^2+\Lambda$,
$\xi_2=(m+\Lambda + \mu)[\Delta^2 - (m+\Lambda)^2 + \mu^2+\Lambda]$, $\xi_3=\Delta[-\Delta^2+(m+\Lambda)^2-\mu^2+\Lambda]$, and $\xi_4=2 \Delta (m+\Lambda+\mu)$ are the parameters shared by the two MESs. Here, $\Lambda=\lambda^2 - k_\parallel^2$ and it satisfies relationship $[- \Delta^2+(m+\Lambda)^2 - \mu^2 -\Lambda]^2=4\Delta^2\Lambda$.
The details for the analytic derivation are shown in Appendix \ref{appendixC}.
Notably, the emergence of MCSs corresponds to the orthogonal conditions $\left<\xi_{u}|\xi_{l}\right>=0$ for the two MESs in the upper and lower layers~\cite{Miao2024}.

\begin{figure}
	\centering
	\includegraphics[width=\columnwidth,angle=0]{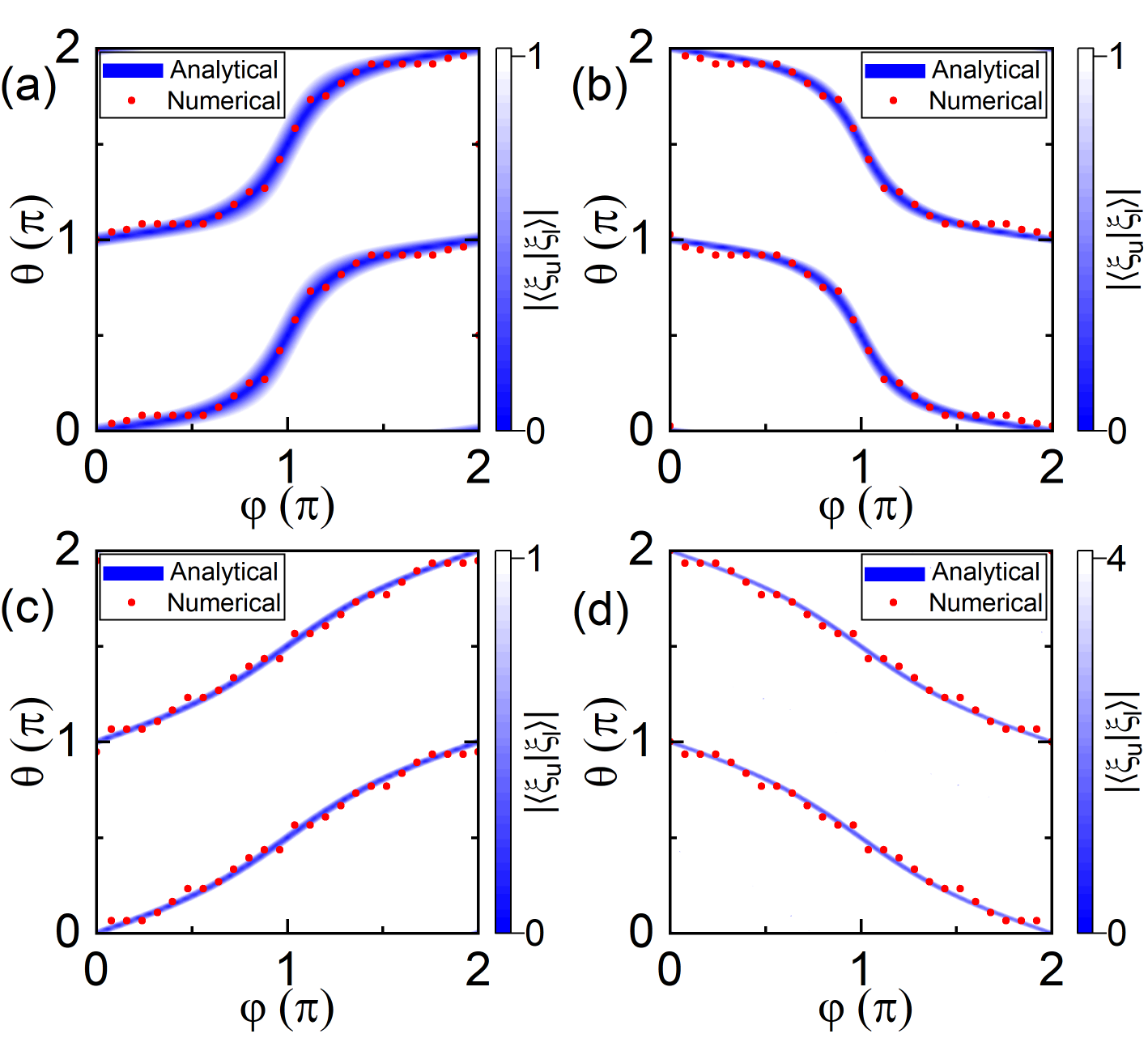}
	\caption{The modules of the inner product of two MES spinors $\left|\left<\xi_{u}|\xi_{l}\right>\right|$ as a function of $\theta$ and $\varphi$. The blue area corresponds to an inner product of zero, where the analytically obtained MES spinors are orthogonal. The red scatter plot numerically shows the angle $\theta_M$ at the MCSs location as a function of $\varphi$. The chemical potential is chosen as $\mu=0.3$ in (a), $\mu=-0.3$ in (b), $\mu=1.5$ in (c), and $\mu=-1.5$ in (d). The other parameters are the same as those in Fig. \ref{fig2}(b).}
	\label{fig3}
\end{figure}

In order to visualize the orthogonal position of the MESs at different superconducting phase biases, we plot the modules of the inner product of two MESs $\left|\left<\xi_{u}|\xi_{l}\right>\right|$ as a function of $\theta$ and $\varphi$ in Fig. \ref{fig3}. We focus on $\left|\left<\xi_{u}|\xi_{l}\right>\right|=0$, highlighted in blue. For $\varphi=0$, two MESs orthogonal at $\theta=0(2\pi)$ and $\theta=\pi$, regardless of the value of $\mu$ [see Fig. \ref{fig3}].
As $\varphi$ changes from $0$ to $2\pi$, the values of $\theta$ corresponding to
$\left|\left<\xi_{u}|\xi_{l}\right>\right|=0$ for the positive (negative) $\mu$
increase (decrease) from $0 (\pi)$ to $\pi(0)$ and from $\pi(2\pi)$ to $2\pi(\pi)$ [Figs. \ref{fig3}].
For $\mu=0.3$ and $-0.3$, it changes slowly near $\varphi = 0$ and $\varphi = 2\pi$, but rapidly near $\varphi = \pi$ [see Figs. \ref{fig3}(a,b)].
But, for $\mu = 1.5$ and $-1.5$, the process is closer to linear [see Figs. \ref{fig3}(c,d)]. These results show that the superconducting phase bias $\varphi$ can effectively manipulate the normal angle $\theta$ that satisfies the two MESs orthogonality conditions. That is, the superconducting phase bias can effectively manipulate the position of the MCSs.

Numerically, the position of the MCSs can be labeled as
$\theta_M$.
Here, $\theta_M=\pm\arctan{\frac{y_2-y_1}{x_2-x_1}}$ is defined as the angle between the lines connecting the two MCSs and the $+x$ axis [see Fig. \ref{fig1}(b)]. $(x_1,y_1)$ and $(x_2,y_2)$ are the lattice coordinates with the highest probability of MCSs.
For comparison, we give the numerical results of $\theta_M$ vs $\varphi$, as represented by discrete red dots in Fig. \ref{fig3}.
As shown in Fig. \ref{fig3}, the red dots are always distributed near blue area. This finding reveals that edge state orthogonality theory effectively predicts the positions of MCSs. The origin of MCSs can be understood as the fact that the MESs remain gapless at a particular angle $\theta=\theta_M$, while they are gapped at other angles $\theta \neq \theta_M$~\cite{Miao2024}. 

\begin{figure}
  \centering
  \includegraphics[width=8.6cm,angle=0]{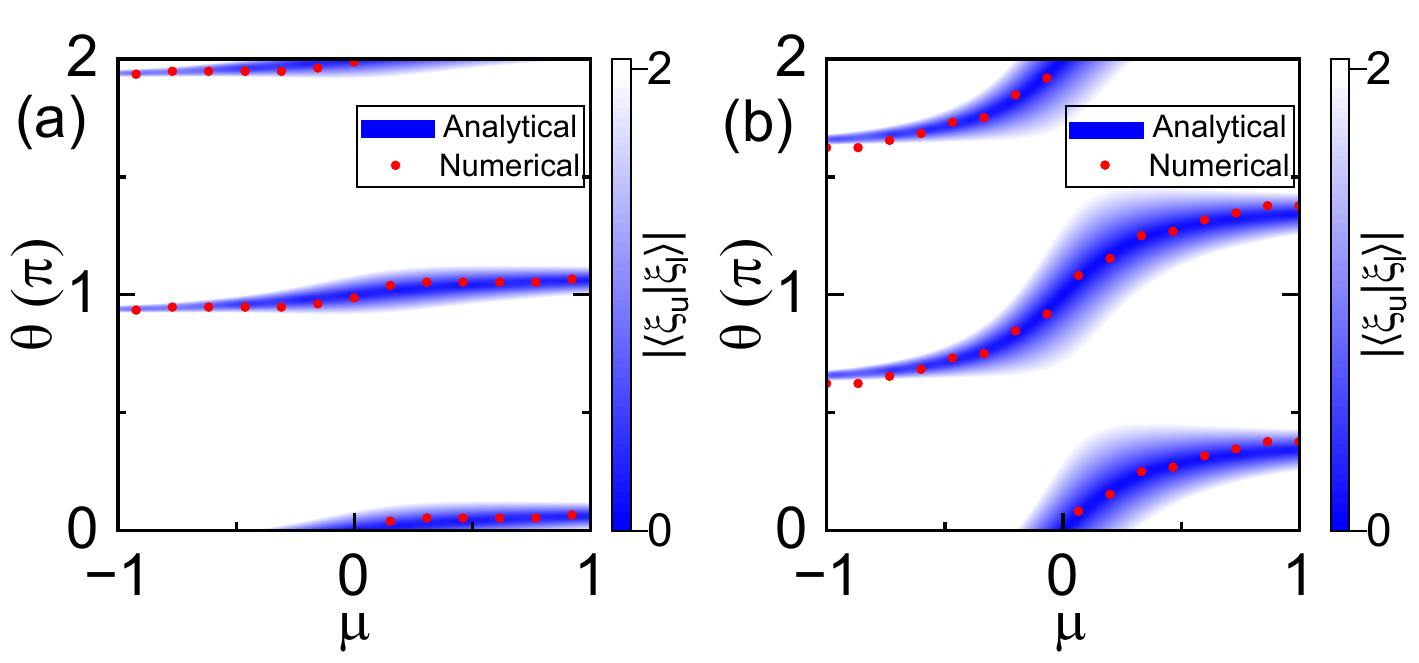}
  \caption{The modules of the inner product of two MESs spinors $\left|\left<\xi_{u}|\xi_{l}\right>\right|$ as a function of $\theta$ and $\mu$. The blue area corresponds to an inner product of zero, where the analytically obtained MES spinors are orthogonal. The red scatter plot numerically shows the angle $\theta_M$ at the MCSs location as a function of the chemical potential $\mu$. The superconducting phase bias is chosen as $\varphi=0.2\pi$ in (a) and $\varphi=0.8\pi$ in (b). The other parameters are set to $m=-0.5$, $A=1$, $B=1$, $\Delta=1$, and $R=25a$.}
  \label{figS2}
\end{figure}
Furthermore, we explore the role of chemical potentials in manipulating the position of the MCSs. We fix the parameters as $m=-0.5$, $\Delta=1$. Regardless of the value of $\mu$, the upper/lower layer remain in the nontrivial topological phase $\mathcal{N}=+1/\mathcal{N}=-1$ without interlayer coupling, as the condition $m^2<\Delta^2+\mu^{2}$ is satisfied.

In the following, we illustrate the role of the chemical potentials in manipulating the MCSs position from both edge state orthogonal analysis and numerical analysis.
In Fig. \ref{figS2}, we plot the modules of the inner product of two MESs spinors $\left|\left<\xi_{u}|\xi_{l}\right>\right|$ as a function of $\theta$ and $\mu$. We focus on $\left|\left<\xi_{u}|\xi_{l}\right>\right|=0$, as illustrated in blue. 
For $\varphi=0.2\pi$, two MESs are orthogonal near $\theta=0(2\pi)$ and $\theta=\pi$, regardless of the value of $\mu$, as shown in Fig. \ref{figS2}(a). In this case, the effect of the chemical potential on the orthogonal angle $\theta$ between two MESs is slight.
While for $\varphi=0.8\pi$, the orthogonal angle between two MESs can be changed effectively by $\mu$ [see Fig. \ref{figS2}(b)]. Additionally, it changes slowly near $\mu = \pm 1$, but rapidly near $\mu=0$. 
These results show that the chemical potential also plays a modulatory role in the orthogonal angles $\theta$ under a large superconducting phase bias.
However, when the superconducting phase bias is small, the chemical potential does not have a significant effect on adjusting the orthogonal angle of the two MESs.

Numerically, we comparatively analyze the effect of chemical potential $\mu$ on the angle $\theta_M$ of emergence of MCSs with different superconducting phase biases in Fig. \ref{figS2}. 
As shown in Fig. \ref{figS2}, the discrete red dots are always distributed near blue area, which directly demonstrates the manipulating of MCSs positions by chemical potentials. This further confirms the accuracy of the edge state orthogonality theory in predicting the position of MCSs. 
However, substantial control of the MCSs by the chemical potential requires certain conditions, specifically that the superconducting phase bias is not very small.
Even with a relatively large phase bias, simply adjusting the chemical potential is insufficient to exchange the two MCSs. In contrast, superconducting phase bias $\varphi$ is more advantageous as it enables more effective and precise manipulation of the MCSs, achieving topological braiding.

\section{\label{sec5}Braiding by swapping two of four MCSs}

\begin{figure}
	\centering
	\includegraphics[width=\columnwidth,angle=0]{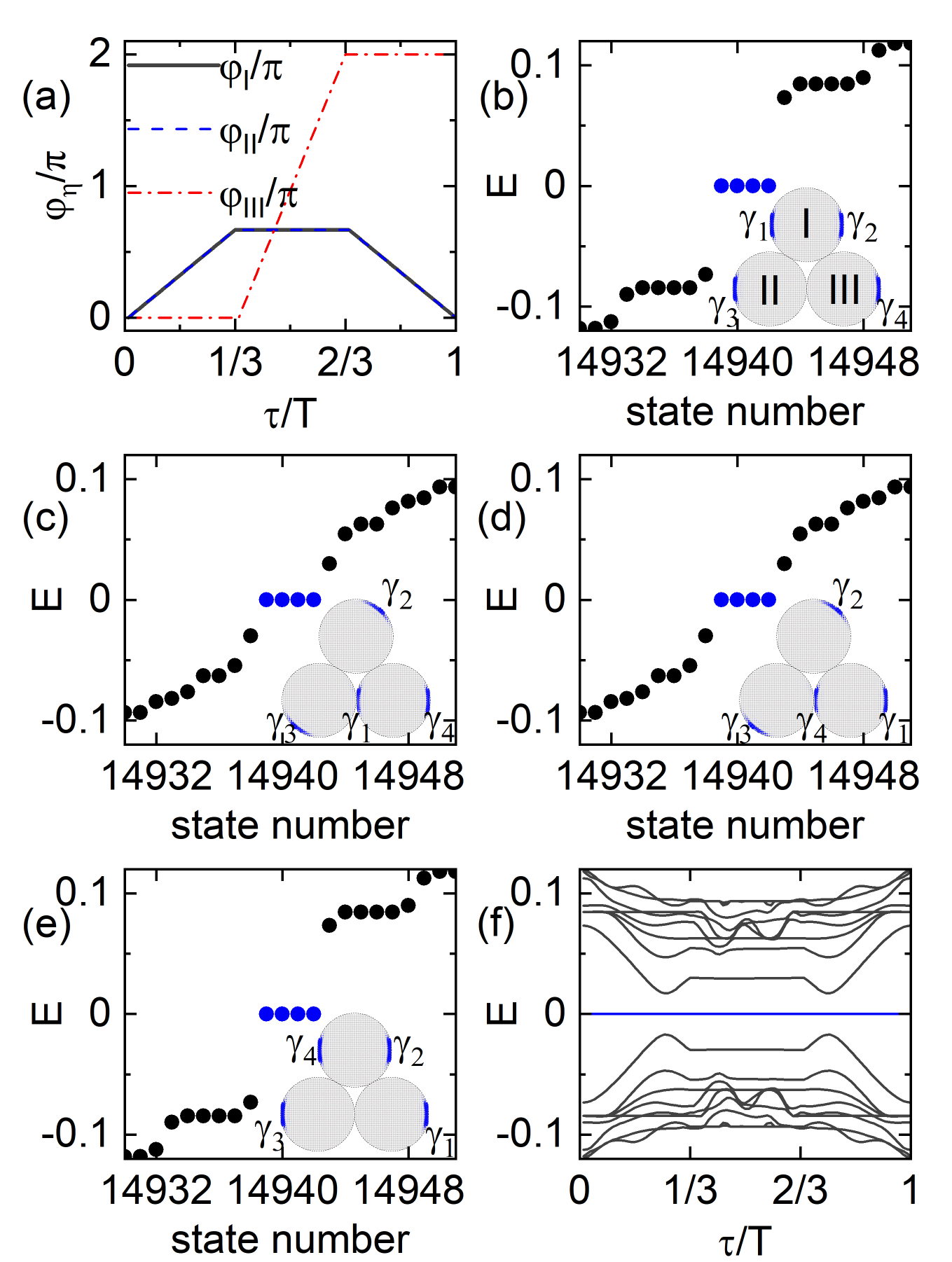}
	\caption{(a) Superconducting phase biases $\varphi_{I,II,III}$ vs time $\tau$ in the hybrid system of three CVJJs.
(b)-(e) Energy levels of the hybrid system with different time $\tau=0$ in (b), $\tau=T/3$ in (c), $\tau=2T/3$ in (d) and $\tau=T$ in (e). The blue dots represent the in-gap MCSs, whose probability distributions are plotted in the inset. (f) Band structure for the hybrid system as a function of time $\tau$. Blue lines indicate the MCSs, which always remain separate from the other states (black lines). We set the parameters as $\mu=1.5$, $R=20a$, and $t_1=0.2$. Other parameters are the same as those in Fig. \ref{fig2}(b).}
	\label{fig4}
\end{figure}

The results above have achieved the exchange of positions between two MCSs induced by $\varphi$ or $\mu$, forming the basis for topological braiding. To further construct a qubit, it is necessary to exchange any two of the four Majorana zero modes.
To achieve this, we design a hybrid system consisting of three CVJJs $\eta=I, II, III$ [see the inset in Fig. \ref{fig4}(b)], in which phase biases $\varphi_\eta$ can be individually controlled. The radius of each CVJJ is set to $R=20a$ and the center coordinates of the three CVJJs are $(0,0)$, $(-R,-\sqrt{3}R)$ and $(R,-\sqrt{3}R)$, respectively.
The Hamiltonian of the hybrid system can be written as $H_{S}=\sum_{\eta}{H_\eta}+H_{c}$.
$H_{\eta}$ is the Hamiltonian of the single junction, which can be described by Eq.~({\ref{eq3}}).
The coupling Hamiltonian $H_{c}$ between the two CVJJs is $H_{c}=t_1\rho_0\tau_z\sigma_0$, where $\rho_{0}$ denotes the $2\times 2$ identity matrix acting on layer degree of freedom and $t_{1}$ is the interjunction coupling strength. The corresponding tight-binding Hamiltonian of the hybrid system is provided in Appendix \ref{appendixA}.

To show the process of the topological braiding, we tune the time-dependent
intensities of superconducting phase bias $\varphi_{\eta}$, as displayed in Fig.~\ref{fig4}(a).
In Figs.~\ref{fig4}(b)-~\ref{fig4}(e), we calculate the energy levels of the hybrid system with different time $\tau$. The braiding protocol takes four steps in $T$ time to spatially swap two MCSs $\gamma_{1}$ and $\gamma_{4}$:
(i) At $\tau=0$, we set $\varphi_{I}=\varphi_{II}=\varphi_{III}=0$.
Figure \ref{fig4}(b) shows that four MCSs (blue dots) are present in the three CVJJs.
Specifically, $\gamma_1$, $\gamma_2$ arise at the left and right corners of CVJJ $I$. While $\gamma_3$ and $\gamma_4$ appear to the left of CVJJ  $II$ and to the right of CVJJ $III$, respectively [see the inset of Fig.~\ref{fig4}(b)]. These four MCSs can be considered as our initial state in the topological braiding processing.
(ii) At $\tau=T/3$, the superconducting phase bias change to $\varphi_{I}=\varphi_{II}=2\pi/3$, $\varphi_{III}=0$. $\gamma_1$ moves to CVJJ $III$, and the remaining three MCSs $\gamma_2$, $\gamma_3$ and $\gamma_4$ are located in the CVJJ $I$, $II$ and $III$, respectively [see Fig.~\ref{fig4}(c)].
(iii) For $\tau \in [T/3, 2T/3]$, we only increase $\varphi_{III}$ from $0$ to $2\pi$, and keep the rest of the phase biases unchanged. It can be seen from Fig.~\ref{fig4}(d) that the position of $\gamma_2$ and $\gamma_3$ remains unchanged, while the positions of $\gamma_1$ and $\gamma_4$ are exchanged in the CVJJ $III$.
(iv) For $\tau \in [2T/3, T]$, we leave the superconducting phase bias of CVJJ $III$ unchanged at $\varphi_{III}=2\pi$ and change the remaining two superconducting phase biases to $\varphi_{I}=0$ and $\varphi_{II}=0$.
One can see from Fig.~\ref{fig4}(e) that  $\gamma_4$, $\gamma_2$ are bounded at the left and right corners of CVJJ $I$, while $\gamma_3$ and $\gamma_1$ appear to the left of CVJJ  $II$ and to the right of CVJJ $III$, respectively.  The spatial positions of $\gamma_1$ and $\gamma_4$ are mutually swapped, and the spatial positions of $\gamma_2$ and $\gamma_3$ remain unchanged [see the inset of Figs.~\ref{fig4}(b) and \ref{fig4}(e)].
In Video of the Supplementary Material~\cite{supp},
we show the animation of the braiding process when the time $\tau$
increases from $0$ to $T$ as shown in Fig.~\ref{fig4}(a).

In order to observe whether the four MCSs are excited throughout the entire braiding process, we plot the energy levels of the hybrid system as a function of the time $\tau$ [see Fig.~\ref{fig4}(f)].
One can see that the MCSs (blue lines) remain stable throughout the variation of $\tau$.
The isolated MCSs can completely prevent mixing with other states (black lines) by an energy gap. Thus, our results validate the stability of the braiding process in the hybrid system.
Furthermore, any two adjacent MCSs of the four MCSs can be swapped
in the hybrid system.

\section{\label{sec6}Summary}

In summary, we propose a scheme for manipulating MCSs by superconducting phase bias in an SC/QAHI/NI/QAHI/SC vertical Josephson junction.
In experiment, the phase bias can be easily controlled by a supercurrent~\cite{Melo2019, Lesser2021} or a Superconducting Quantum Interference Device~\cite{Fagaly2006, Giazotto2010, Noh2021}.
The recent development of microwave pulse generators has been designed for convenient digital control of the superconducting phase \cite{Bao2024}.
It is demonstrated that the position of MCSs in this setup 
can be precisely controlled by the phase bias.
The origin and movement of the MCSs is effectively explained through the orthogonality of edge states.
Furthermore, we propose a protocol to implement the swap of any two MCSs of four MCSs in the hybrid system consisting of three CVJJs.
Importantly, MCSs remain stable throughout the braiding process.
The stability and controllability of MCSs in this system have significant implications for topological quantum computing.

\begin{acknowledgments}
This work was financially supported by the National Key Research and Development Program of China (Grant No. 2024YFA1409002), the National Natural Science Foundation of China (Grants No. 12447147, No. 124B2069, No. 12374034 and No. 12074097), the Natural Science Foundation of Hebei Province (Grant No. A2024205025), the Innovation Program for Quantum Science and Technology (Grant No. 2021ZD0302403), and the China Postdoctoral Science Foundation (Grant No. 2024M760070). We also acknowledge the High-performance Computing Platform of Peking University for providing computational resources.
\end{acknowledgments}
\appendix
\renewcommand{\appendixname}{APPENDIX}

\setcounter{equation}{0}
\renewcommand{\theequation}{A\arabic{equation}}
\section{\label{appendixA} \MakeUppercase{Tight-binging model}}

Since the tight-binding representation is used in our calculations, the continuous Hamiltonian of the single vertical Josephson junction can be mapped onto a tight-binding representation on a two-dimensional square lattice as:
  \begin{align}
  H=&\sum_{\mathbf{i}}^{}{\left[ \psi _{\mathbf{i}}^{\dagger}T_0\psi _{\mathbf{i}}+\left( \psi _{\mathbf{i}}^{\dagger}T_x\psi _{\mathbf{i}+\delta \mathbf{x}}+\psi _{\mathbf{i}}^{\dagger}T_y\psi _{\mathbf{i}+\delta \mathbf{y}} \right) +\text{H.c}. \right]},\nonumber\\
T_0=&\left( m-4B \right) \rho _0\tau _z\sigma _z+\mu \rho _0\tau _z\sigma _y+t\rho _x\tau _0\sigma _0\nonumber\\
&-\Delta \frac{\rho _0+\rho _z}{2}\tau _y\sigma _y-\Delta \cos \varphi \frac{\rho _0-\rho _z}{2}\tau _y\sigma _y\nonumber\\
&+\Delta \sin \varphi \frac{\rho _0-\rho _z}{2}\tau _x\sigma _y,\nonumber \\
T_x=&B\rho _0\tau _0\sigma _z+\frac{A}{2i}\rho _z\tau _0\sigma _x,\quad
T_y=B\rho _0\tau _0\sigma _z+\frac{A}{2i}\rho _0\tau _z\sigma _y,
\label{eqs1}
 \end{align}
where the basis $\psi_{\bf i}^{\dagger}=(c_{{\bf i},\uparrow}^{u\dagger},c_{{\bf i},\downarrow}^{u\dagger},c_{{\bf i},\uparrow}^{u},c_{{\bf i},\downarrow}^{u},c_{{\bf i},\uparrow}^{l\dagger},c_{{\bf i},\downarrow}^{l\dagger},c_{{\bf i},\uparrow}^{l},c_{{\bf i},\downarrow}^{l})$, $c_{{\bf i}, \uparrow / \downarrow}^{u/l\dagger}$ and $c_{{\bf i}, \uparrow / \downarrow}^{u/l}$ are the creation and annihilation operator on site $i$ of the upper/lower layer with spin $\uparrow/\downarrow$.
${\bf i}=(x,y)$ is the coordinates of the size with $x$ and $y$ being integers
and $\delta \mathbf{x}$ $(\delta \mathbf{y})$ is the unit vector along the $x$ ($y$) direction.
$\rho_{0}$ and $\rho_{x,y,z}$ are the $ 2\times 2$ unit matrix and the Pauli matrices acting on layer degree of freedom.
The lattice constant has been set to $a=1$ here.

\begin{figure}
  \centering
\includegraphics[width=0.9\columnwidth,angle=0]{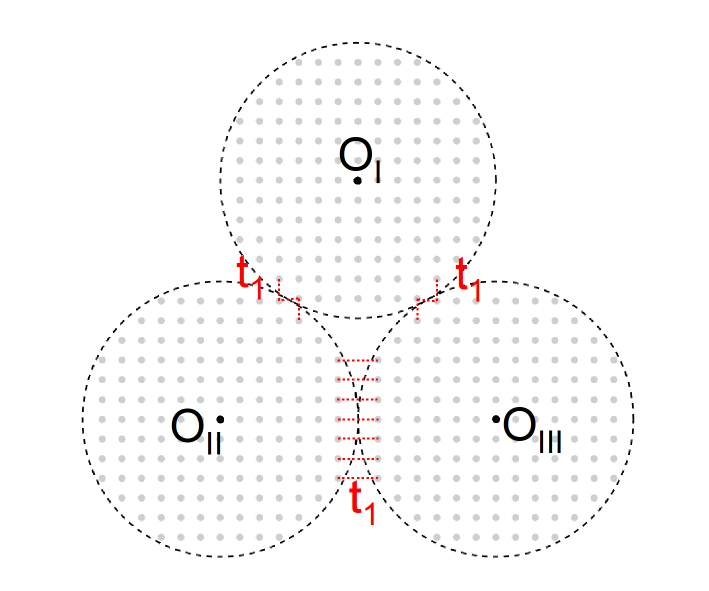}
  \caption{Lattice configuration of the hybrid system consisting of three CVJJs. The gray dots are the lattice sites, the black dashed circles are the CVJJ boundaries, and the red dotted lines represent interjunction hopping with strength $t_1$ between two CVJJs. $O_{I}=(0,0)$, $O_{II}=(-7a,7\sqrt{3}a)$, and $O_{III}=(7a,7\sqrt{3}a)$ are the coordinates of the three CVJJs' centers (black dots) with the radius $R=7a$.}
  \label{figS1}
\end{figure}

For clarity, we choose the radius $R = 7a$ as an example to show the interjunction coupled lattice sites of the hybrid system consisting of three CVJJs in Fig. \ref{figS1}. 
The center coordinates of the three CVJJs are $O_{I}=(0,0)$, $O_{II}=(-7a,-7\sqrt{3}a)$ and $O_{III}=(7,-7\sqrt{3}a)$, respectively.
We consider that interjunction coupling $t_1$ exists only between lattice sites at the boundaries of the CVJJs. According to the lattice positions shown in Fig. \ref{figS1}, the shortest lattice distance between CVJJs $II$ and $III$ is $d_{II,III}=2a$, so
we set that coupling $t_1$ occurs between lattice sites separated by a distance of $d_{II,III}=2a$. 
However, the distance between CVJJs $I$ and $II$($III$) is smaller. Here, we assume that coupling $t_1$ exists between lattice sites separated by a distance of $d_{I,II(III)}<\sqrt{2}a$. In fact, choosing $\sqrt{2}a$ or another value smaller than $2a$ does not affect the results.
The tight-binding Hamiltonian for a single CVJJ is described by Eq. (\ref{eqs1}), and the coupling Hamiltonian between the CVJJs is chosen as:
\begin{align}
H_{c}=&t_1 \rho_0\tau_z\sigma_0 \left( \sum_{j_{I},j_{II}}^{d_{I,II}<\sqrt{2}}{\psi_{j_{I}}^{\dagger}\psi_{j_{II}}}+\sum_{j_{I},j_{III}}^{d_{I,III}<\sqrt{2}}{\psi_{j_{I}}^{\dagger}\psi_{j_{III}}}\right. \nonumber \\
&+ \left. \sum_{j_{II},j_{III}}^{d_{II,III}= 2}{\psi_{j_{II}}^{\dagger}\psi_{j_{III}}}\right)+\text{H.c.}.
\label{eqs2}
\end{align}
Here, the lattice distance between the lattice sites $j_\eta$ of the CVJJ $\eta$ and $j_{\eta^{\prime}}$ of the CVJJ $\eta^{\prime}$ can be numerically calculated as 
$d_{\eta,\eta^{\prime}}=\sqrt{(x_{j_{\eta}}-x_{j_{\eta^{\prime}}})^2+(y_{j_{\eta}}-y_{j_{\eta^{\prime}}})^2}$.

\setcounter{equation}{0}
\renewcommand{\theequation}{B\arabic{equation}}
\section{\label{appendixB} \MakeUppercase{Manipulating MCSs in the elliptic vertical Josephson junctions}}
\begin{figure*}
	\centering
\includegraphics[width=1.9\columnwidth,angle=0]{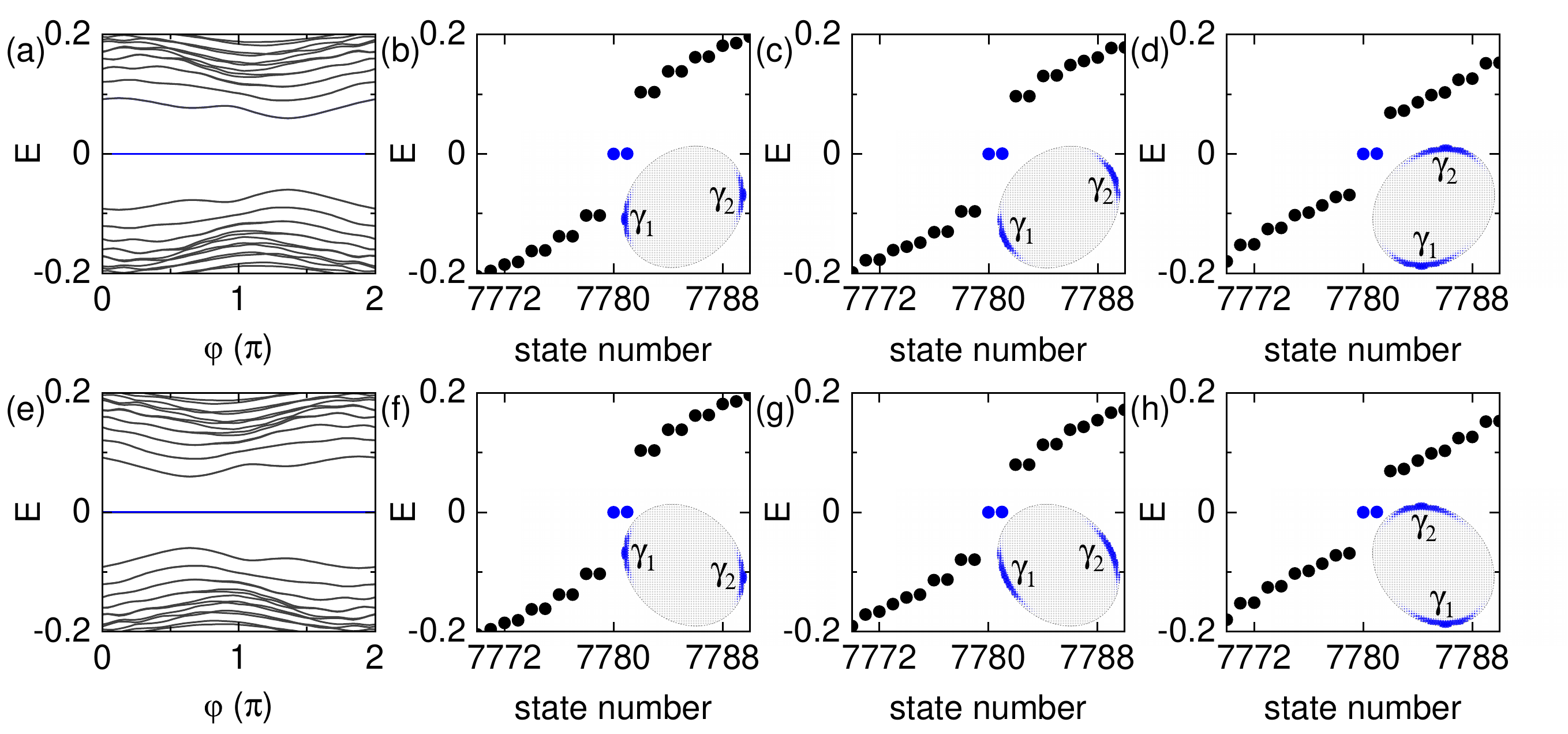}
	\caption{
    (a,e) Energy spectrum of the elliptic vertical Josephson junction as a function of the superconducting phase bias $\varphi$. Blue lines represent the zero-energy Majorana bands, which always remain separate from the other bands (black lines). (b–d, f–h) Spatial probability distributions of the in-gap MCSs with different superconducting phase biases $\varphi = 0$ for (b,f), $\varphi = 0.5\pi$ for (c,g), and $\varphi = \pi$ for (d,h). The blue dots represent the in-gap MCSs, the probability distributions of the MCSs are plotted in the insets. The angle between the major axis of the ellipse and the $+x$-direction is set to $\alpha=\pi/4$ for (a-d) and $\alpha=3\pi/4$ for (e-h). The lengths of the semi-major axis and semi-minor axis of the ellipse are fixed as $R_{a}=27.5a$ and $R_b=22.5a$, respectively. The other parameters remain the same as those in Fig. \ref{fig2}}.
	\label{figS3}
\end{figure*}
To illustrate that the manipulation of MCSs is insensitive to the geometric shape of the junction, we plot the energy levels of the elliptic vertical Josephson junctions and the spatial distributions of the MCSs in Fig. \ref{figS3}. Specifically, the elliptic system is defined on a square lattice centered at the origin $(0,0)$, where each lattice site is labeled by coordinates $(x,y)$. The sites belonging to the elliptic region satisfy the inequality:
\begin{align}
\frac{\left( x \cos \alpha + y \sin \alpha \right)^2}{R_a^2} + \frac{\left( -x \sin \alpha + y \cos \alpha \right)^2}{R_b^2} < 1,
\label{eqs16}
\end{align}
where $\alpha$ is the angle between the major axis of the ellipse and the $+x$-direction. The parameters $R_a = 27.5a$ and $R_b = 22.5a$ denote the lengths of the semi-major and semi-minor axes, respectively. In the calculations, we fix $R_{a}=27.5a$, $R_{b}=22.5a$, and set $\alpha=\pi/4$ in Figs. \ref{figS3}(a-d) and $\alpha=\pi/4$ in Figs. \ref{figS3}(e-h). All other parameters remain the same as those in Fig. \ref{fig2}.

Figures~\ref{figS3}(a) and \ref{figS3}(e) display the energy levels of the elliptic vertical Josephson junction as a function of the superconducting phase bias $\varphi$ with different angles $\alpha=\pi/4$ and $\alpha=3\pi/4$, respectively. In both cases, the zero-energy in-gap states (blue lines) remain well separated from the other bands (black lines) throughout the full evolution of $\varphi$. This confirms that the MCSs persist and remain topologically protected even if the geometric shape of the system is elliptical with different angles.

To further explore the spatial behavior of the MCSs, we plot their probability distributions at $\varphi = 0$, $0.5\pi$, and $\pi$, as shown in Figs.~\ref{figS3}(b–d) for $\alpha = \pi/4$ and Figs.~\ref{figS3}(f–h) for $\alpha = 3\pi/4$. In both configurations, two MCSs remain well localized at specific corners along the elliptical boundary. As the phase bias $\varphi$ increases, their positions shift smoothly and predictably along the edge. Importantly, the positions of the MCSs are determined solely by the local edge angle $\theta$, which satisfies the edge states orthogonality condition, and are independent of the geometric shape of the junction.

These results clearly demonstrate that the phase-controlled manipulation of MCSs is robust against variations in the junction shape. The emergence and exchange of MCSs are governed by the edge angle $\theta$, which remains fixed once the system parameters are specified. Therefore, the superconducting phase bias controlled braiding protocol proposed in the main text remains applicable even in anisotropic geometries, reinforcing the generality and flexibility of the scheme.

\begin{widetext}
\setcounter{equation}{0}
\renewcommand{\theequation}{C\arabic{equation}}
\section{\label{appendixC} \MakeUppercase{The effective edge theory and orthogonal conditions}}

We now take the lower MES as an example to derive the effective edge theory of the low-energy Hamiltonian:
\begin{align}
  H_{\mathbf{k}}^{l}=&(m-B\mathbf{k}^2)\tau_{z}\sigma_{z}-Ak_{x}\tau_{0}\sigma_{x}+Ak_{y}\tau_{z}\sigma_{y} +\mu\tau_{z}\sigma_{0}-\Delta\cos{\varphi}\tau_{y}\sigma_{y}+\Delta\sin{\varphi}\tau_{x}\sigma_{y}.
  \label{eqs3}
 \end{align}
We consider a circular-shaped boundary, the normal direction of the boundary tangent for arbitrary angle $\theta$ is $\hat{x}_{\bot}= (\cos \theta, \sin \theta)$. 
Equation (\ref{eqs3}) can be rewritten in terms of $k_x=\cos{\theta}k_{\bot}+\sin{\theta}k_{\parallel}$ and $k_y=\sin{\theta}k_{\bot}-\cos{\theta}k_{\parallel}$ as:
\begin{align}
	H(k_{\bot},k_{\parallel})=&(m-Bk^2)\tau_{z}\sigma_{z}+A(-\cos{\theta}\tau_{0}\sigma_{x}+\sin{\theta}\tau_{z}\sigma_{y})k_{\bot}+A(-\sin{\theta}\tau_{0}\sigma_{x}-\cos{\theta}\tau_{z}\sigma_{y})k_{\parallel} +\mu\tau_{z}\sigma_{0}\nonumber \\&-\Delta\cos{\varphi}\tau_{y}\sigma_{y}+\Delta\sin{\varphi}\tau_{x}\sigma_{y}.
	\label{eqs4}
\end{align}
To obtain the corresponding low-energy Hamiltonian which describes the MESs, we perform the replacement $k_{\bot} \rightarrow -i \partial_{\bot}$ \cite{Tan2022, Pan2024, Zhang2020}: 
\begin{align}
	H(\partial_{\bot},k_{\parallel})=&(m+B\partial_{\bot}^{2}-Bk_{\parallel}^2)\tau_{z}\sigma_{z}-iA\partial_{\bot}(-\cos{\theta}\tau_{0}\sigma_{x}+\sin{\theta}\tau_{z}\sigma_{y})+Ak_{\parallel}(-\sin{\theta}\tau_{0}\sigma_{x}-\cos{\theta}\tau_{z}\sigma_{y}) +\mu\tau_{z}\sigma_{0}\nonumber \\&-\Delta\cos{\varphi}\tau_{y}\sigma_{y}+\Delta\sin{\varphi}\tau_{x}\sigma_{y}.
	\label{eqs5}
\end{align}

Next, we assume an ansatz for the edge state wave function at $\theta$ as $\Psi_{l}(x_{\bot})=e^{\lambda x_{\bot}}e^{i k_{\parallel} x_{\parallel}} \left|\xi_{l}\right>$. Plugging this ansatz into Eq. (\ref{eqs5}), we obtain the eigen equation with eigenenergy $E=0$:

\begin{align}
		\left[\begin{matrix}
		(m+B\lambda^{2}-Bk_{\parallel}^2)+\mu & iA e^{i\theta} (\lambda+k_{\parallel}) & 0 & \Delta e^{-i\varphi} \\
		iA e^{-i\theta} (\lambda-k_{\parallel})& -(m+B\lambda^{2}-Bk_{\parallel}^2)+\mu & -\Delta e^{-i\varphi} & 0 \\
		0 & -\Delta e^{i\varphi} & -(m+B\lambda^{2}-Bk_{\parallel}^2)-\mu&  iA e^{-i\theta} (\lambda-k_{\parallel})\\ 
		\Delta e^{i\varphi} & 0 & iA e^{i\theta} (\lambda+k_{\parallel})& (m+B\lambda^{2}-Bk_{\parallel}^2)-\mu  
	\end{matrix}\right]
    \left[\begin{matrix}
		\xi_a \\
		\xi_b \\
		\xi_c \\
		\xi_d
	\end{matrix}\right]=0.
	\label{eqs6}
\end{align}

For simplicity, we set $A = B = 1$. A nontrivial solution of $(\xi_a,\xi_b,\xi_c,\xi_d)^T$ to Eq. (\ref{eqs6}) yields:
\begin{align}
[(m+\lambda^2 - k_\parallel^2)^2 - \Delta^2- \mu^2 -\lambda^2 + k_\parallel^2]^2=4\Delta^2(\lambda^2 - k_\parallel^2).
	\label{eqs7}
\end{align}
We rewrite the above equation using $\Lambda=\lambda^2-k_\parallel^2$ as:
\begin{align}
[(m+\Lambda)^2 - \Delta^2- \mu^2 -\Lambda]^2=4\Delta^2\Lambda.
	\label{eqs8}
\end{align}
Solving this, we find eight solutions of $\lambda$ as
$\lambda_{1/2/3/4/5/6/7/8}=\pm \sqrt{\Lambda_{1/2/3/4}+k_\parallel^2}$.
Each $\lambda$ corresponds to a spinor solution of $(\xi_a,\xi_b,\xi_c,\xi_d)^T$. 
Considering a half-infinite sample area $x_{\bot} < 0$ with the boundary conditions $\Psi_l(x_\bot=-\infty)=e^{\lambda (-\infty)}e^{i k_{\parallel} x_\parallel} \xi_l=0$, so we keep only the four solutions with $Re[\lambda_{\beta}]>0$ ($\beta=1,2,3,4$). The corresponding four spinors solutions $\xi^{\lambda_{\beta}}=(\xi_a^{\lambda_{\beta}},\xi_b^{\lambda_{\beta}},\xi_c^{\lambda_{\beta}},\xi_d^{\lambda_{\beta}})^T$ are preserved. Assuming $\xi_a^{\lambda_{\beta}}=1$, a special solution of the spinor is obtained in the form of:

\begin{align}
\left|\xi_l\right>=\left[\begin{matrix}
		1 \\
		(m+\Lambda+ \mu)(\Delta^2 - (m+\Lambda)^2 + \mu^2+\Lambda)/[-i  e^{i\theta} (\lambda+k_{\parallel}) (-\Delta^2-(m+\Lambda)^2+\mu^2+\Lambda)]\\
	 \Delta(-\Delta^2+(m+\Lambda)^2-\mu^2+\Lambda)/[-i  e^{i(\theta-\varphi)} (\lambda+k_{\parallel}) (-\Delta^2-(m+\Lambda)^2+\mu^2+\Lambda)] \\
		2 \Delta (m+\Lambda+\mu)/[e^{-i\varphi}(-\Delta^2-(m+\Lambda)^2+\mu^2+\Lambda)]
	\end{matrix}\right].
    \label{eqs9}
\end{align}

We impose the open boundary conditions $\Psi(x_\bot=0)=e^{i k_{\parallel} x_\parallel} \xi=0$ to the wave function, leading to a homogeneous linear equation $C_1\xi^{\lambda_{1}}+C_2\xi^{\lambda_{2}}+C_3\xi^{\lambda_{3}}+C_4\xi^{\lambda_{4}}=0$, 
where ${C_1,C_2,C_3,C_4}$ are not all zero coefficients. The existence of such a nontrivial solution implies that the determinant $\left|\xi^{\lambda_{1}}, \xi^{\lambda_{2}}, \xi^{\lambda_{3}}, \xi^{\lambda_{4}}\right|$ equal to 0.
This, together with the expressions of $\lambda_{\beta}$ in Eq. (\ref{eqs7}), we can obtain $k_\parallel=0$. Plugging this into Eq. (\ref{eqs9}) and simplifying gives:
\begin{align}
   \left|\xi_l\right>=\left[\begin{matrix}
		-\Delta^2-(m+\Lambda)^2+\mu^2+\Lambda \\
		(m+\Lambda + \mu)[\Delta^2 - (m+\Lambda)^2 + \mu^2+\Lambda]/[-i  e^{i\theta} (\lambda+k_\parallel) ]\\
	\Delta[-\Delta^2+(m+\Lambda)^2-\mu^2+\Lambda]/[-i  e^{i(\theta-\varphi)} (\lambda+k_\parallel) ] \\
		2 \Delta (m+\Lambda+\mu)/[e^{-i\varphi}]
	\end{matrix}\right].
    \label{eqs10}
\end{align}
Similarly, the spinor for the Chiral MES in the upper layer is:
\begin{align}
   \left|\xi_u\right>=\left[\begin{matrix}
		-\Delta^2-(m+\Lambda)^2+\mu^2+\Lambda \\
		(m+\Lambda + \mu)[\Delta^2 - (m+\Lambda)^2 + \mu^2+\Lambda]/[i  e^{-i\theta} (\lambda-k_\parallel)]\\
	\Delta[-\Delta^2+(m+\Lambda)^2-\mu^2+\Lambda]/[i  e^{-i\theta} (\lambda-k_\parallel) ] \\
		2 \Delta (m+\Lambda+\mu)
	\end{matrix}\right].
    \label{eqs11}
\end{align}
By comparing the forms of the two spinors, we observe that they share common parameters:
\begin{align}
\xi_1&=-\Delta^2-(m+\Lambda)^2+\mu^2+\Lambda,\quad\quad\quad
\xi_2=(m+\Lambda + \mu)[\Delta^2 - (m+\Lambda)^2 + \mu^2+\Lambda], \nonumber \\
\xi_3&=\Delta[-\Delta^2+(m+\Lambda)^2-\mu^2+\Lambda],\quad\quad
\xi_4=2 \Delta (m+\Lambda+\mu).
    \label{eqs12}
\end{align}
Therefore, the two spinors can be simplified as:
\begin{align}
\small
   \left|\xi_u\right>=\left[\begin{matrix}
		\xi_1 \\
		\xi_2/[i  e^{-i\theta} (\lambda-k_\parallel) ]\\
	\xi_3/[i  e^{-i\theta} (\lambda-k_\parallel) ] \\
		\xi_4
	\end{matrix}\right],\quad
   \left|\xi_l\right>=\left[\begin{matrix}
		\xi_1 \\
		\xi_2/[-i  e^{i\theta} (\lambda+k_\parallel) ]\\
	\xi_3/[-i  e^{i(\theta-\varphi)} (\lambda+k_\parallel) ] \\
		\xi_4/[e^{-i\varphi}]
	\end{matrix}\right].
    \label{eqs13}
\end{align}

It is noteworthy that the occurrence of MCSs corresponds to the orthogonal conditions $\left<\xi_u|\xi_l\right>=0$ of the MESs in the upper and lower layers. 
We analyze analytically the case where the $\Lambda$ is a purely real solution, and the condition for the orthogonality of the two MESs reduces to:
\begin{align}
\left( m+\Lambda-\mu \right) \left[ 1-e^{-i\left( 2\theta -\varphi \right)} \right] +\left( m+\Lambda+\mu \right) \left[ -e^{-2i\theta}+e^{i\varphi} \right] =0.
\label{eqs14}
\end{align}

For $\mu=0$, the edge states orthogonality condition reduces to 
\begin{align}
1-e^{-i\left( 2\theta -\varphi \right)}-e^{-2i\theta}+e^{i\varphi}=0.
\label{eqs15}
\end{align}

It is obvious that Eq. (\ref{eqs15}) holds at any edge angle $\theta$ with $\varphi=\pi$. 
But as long as $\mu \neq 0$, $\theta=\theta_M$ satisfying the orthogonality condition can be adjusted by parameters such as the superconducting phase bias $\varphi$ and chemical potential $\mu$.
The MESs of the upper and lower layers remain orthogonal only
at edge $\theta=\theta_M$, which is the basis for the existence of MCSs.

\end{widetext}

\end{document}